\documentclass[12pt]{iopart}
\usepackage{graphicx}
\usepackage{dcolumn}
\usepackage{bm}
\usepackage{setspace}
\usepackage{psfrag}
\usepackage{subfigure}
\usepackage{amsfonts}
\usepackage{braket}
\usepackage{color}
\usepackage{pdfpages}
\newcommand{\be}{\begin{equation}}
\newcommand{\ee}{\end{equation}}
\newcommand{\bea}{\begin{eqnarray}}
\newcommand{\eea}{\end{eqnarray}}

\newcommand{\bef}{\begin{figure}}
\newcommand{\enf}{\end{figure}}
\newcommand{\p}{\mathbf{p}}
\newcommand{\rmbf}{\mathbf{r}}

\newcommand{\A}{\mathbf{A}}
\newcommand{\D}{\mathbf{D}}
\let\d\undefined
\newcommand{\d}{\mathbf{d}}

\begin{document}

\title[Multidimensional high harmonic spectroscopy]{Multidimensional high harmonic spectroscopy: A semi-classical perspective on measuring
multielectron rearrangement upon ionization}
\author{Valeria Serbinenko and Olga Smirnova}
\address{Max Born Institute, Max-Born-Str. 2A, 12489, Berlin, Germany}
\eads{\mailto{serbinen@mbi-berlin.de}, \mailto{olga.smirnova@mbi-berlin.de}}

\begin{abstract}
High harmonic spectroscopy has the potential to combine attosecond temporal with sub-Angstrom
spatial resolution of the early nuclear  and multielectron dynamics
in molecules. It involves  strong field ionization  of the molecule by the IR laser field  followed by time-delayed recombination of the removed electron with the molecular ion.
The time-delay is controlled on the attosecond time scale by the oscillation of the IR field and is mapped into the harmonic number, providing a movie of molecular dynamics between ionization and recombination.
One of the challenges in the analysis of high harmonic signal stems from the fact that the complex dynamics of both
ionization and recombination  with their multiple observables are entangled in the harmonic signal. Disentangling this information requires
multidimensional approach, capable of mapping ionization and recombination dynamics into different independent
parameters. We suggest multidimensional high harmonic spectroscopy as a tool for characterizing of ionization and recombination processes separately allowing for simultaneous detection of both the ionization delays and sub-cycle ionization rates. Our method extends the capability of the two dimensional (2D) set-up suggested recently by Shafir et al on reconstructing ionization delays, while  keeping the reconstruction procedure as simple as in the original proposal. The scheme is based on the optimization of the high harmonic signal in orthogonally polarized strong fundamental and relatively weak multicolour control fields.
\end{abstract}
\submitto{\jpb}

\maketitle


\section{Introduction}
The possibility of correlation driven sub-femtosecond hole dynamics upon photoionization of molecules
has been a subject of intense theoretical study in the past decade
  \cite{Cederbaum,levine,kuleff1,kuleff2,kuleff3}.
  Attosecond technology is now making it possible to time-resolve  ultrafast processes of electron
 removal \cite{Schultze,Eckle,Pfeiffer,Klunder,Shafir} and subsequent  hole motion \cite{smirnova09,Haessler,Goulielmakis,Mairesse} in atoms and molecules.
 One way to record the hole dynamics is offered by
high harmonic spectroscopy, which  has the potential to combine attosecond temporal with sub-Angstrom
spatial resolution of early nuclear \cite{Baker} and multielectron dynamics \cite{smirnova09,Haessler,Mairesse}
in molecules. It exploits a build-in pump-probe process driven by infrared (IR) laser field and repeated
every half cycle of its oscillation. The pump step is a strong field ionization, which removes an
electron and creates a hole in the system. The oscillating electron returns to the core and probes
it via radiative recombination. High harmonics of driving frequency emitted during recombination record
the shape, position and momentum of the hole for different delays between ionization and recombination \cite{smirnova09}.
Thus, high harmonic spectroscopy records hole dynamics on the attosecond time scale.

 Quantum mechanically, the hole  dynamics is encoded in complex ionization amplitudes reflecting the quantum
  state of the electron and  the ion during and after ionization.
 The phase of the complex ionization amplitude is accumulated not only  due to
 the electron interaction with the core potential and electron-electron correlations,
  but -- thanks to the presence of  the IR field -- also due to the induced polarization and excitations prior to ionization.
 These interactions also affect the magnitude of the ionization amplitude modifying
  the ionization probability. 

    One of the challenges in the analysis of high harmonic signal stems from the fact that the complex dynamics of both
ionization and recombination steps, with their multiple observables, are entangled in the harmonic signal. Disentangling these steps requires
multidimensional approach, capable of mapping ionization and recombination dynamics into different independent
parameters.

Although core rearrangement is a quantum process, which involves multiple ionizaton channels and therefore electron-core entanglement,
 an important insight into its dynamics can come from the semiclassical analysis of the motion of the ejected electron,
  described by the concept of quantum orbits \cite{pascal}.
 Quantum orbits are trajectories evolving in complex time, and characterized by complex velocities and displacements \footnote{  The language of complex trajectories arises in high harmonic generation from applying the saddle-point approximation \cite{lewenstein,chapter} for computing induced dipole moment. Within this perspective, the strong field ionization is interpreted as tunnelling along the complex trajectory through the barrier created by the strong laser field and the Coulomb potential of the molecule \cite{lewenstein,Keldysh,PPT,chapter}. While originally introduced within the strong field approximation (SFA), which neglects the effects of the core potential on the liberated electron, quantum trajectories are valid outside the SFA, see e.g. \cite{lisa1,lisa2,jivesh}.}.
Core rearrangement leaves its imprint on the dynamics
of ejected electron affecting  the time when the electron starts its motion in the continuum, its initial velocity and  angular momentum.
Thus, it must affect the quantum orbits.
For example, the phase of the  ionization amplitude is mapped into the ionization time - the time when the electron appears in the continuum \cite{Larmor}.
 The magnitude of  the ionization amplitude maps into the so-called imaginary ionization time - another parameter of the quantum orbit \cite{Keldysh,PPT,chapter}.
 The initial angular momentum or initial velocity  may be affected due to  inelastic electron-core interactions \cite{walters,lisa2}
 during ionization, involving energy and momentum transfer \cite{pisanty} between the electron and the core further modifying the quantum orbit.

\begin{figure}
\label{figure1}
\begin{center}
\includegraphics[width=11cm]{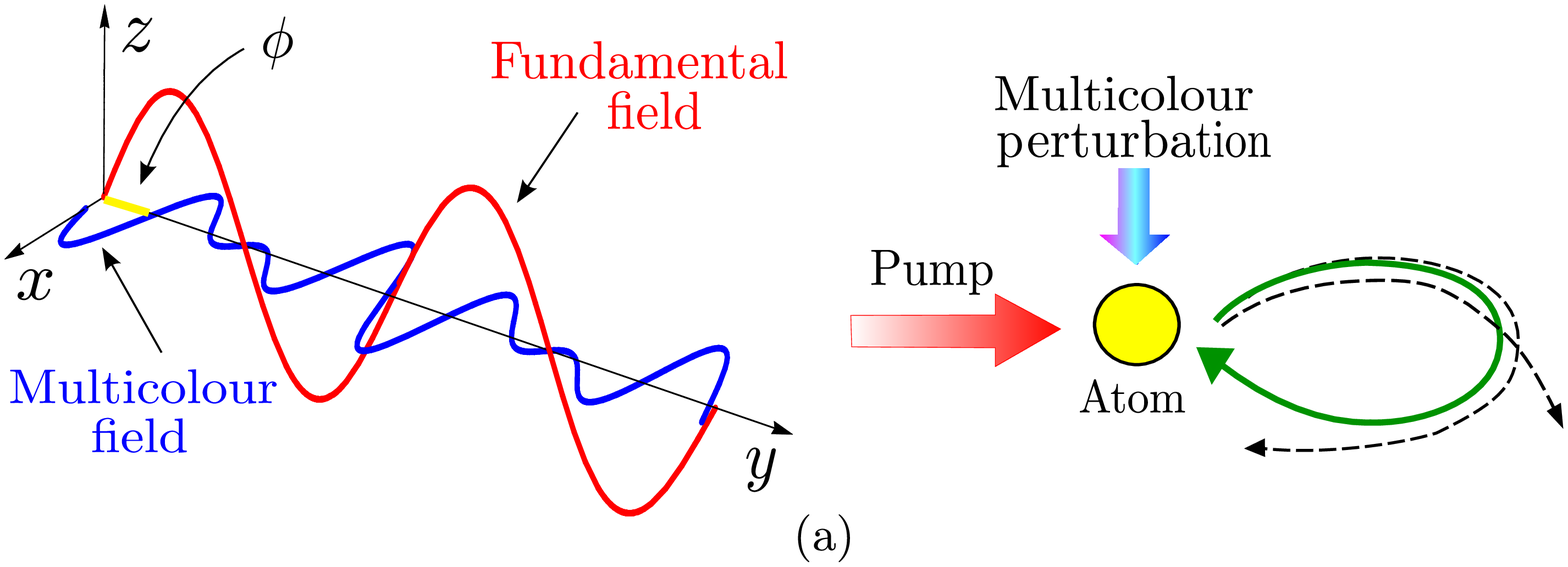}
\includegraphics[width=4cm]{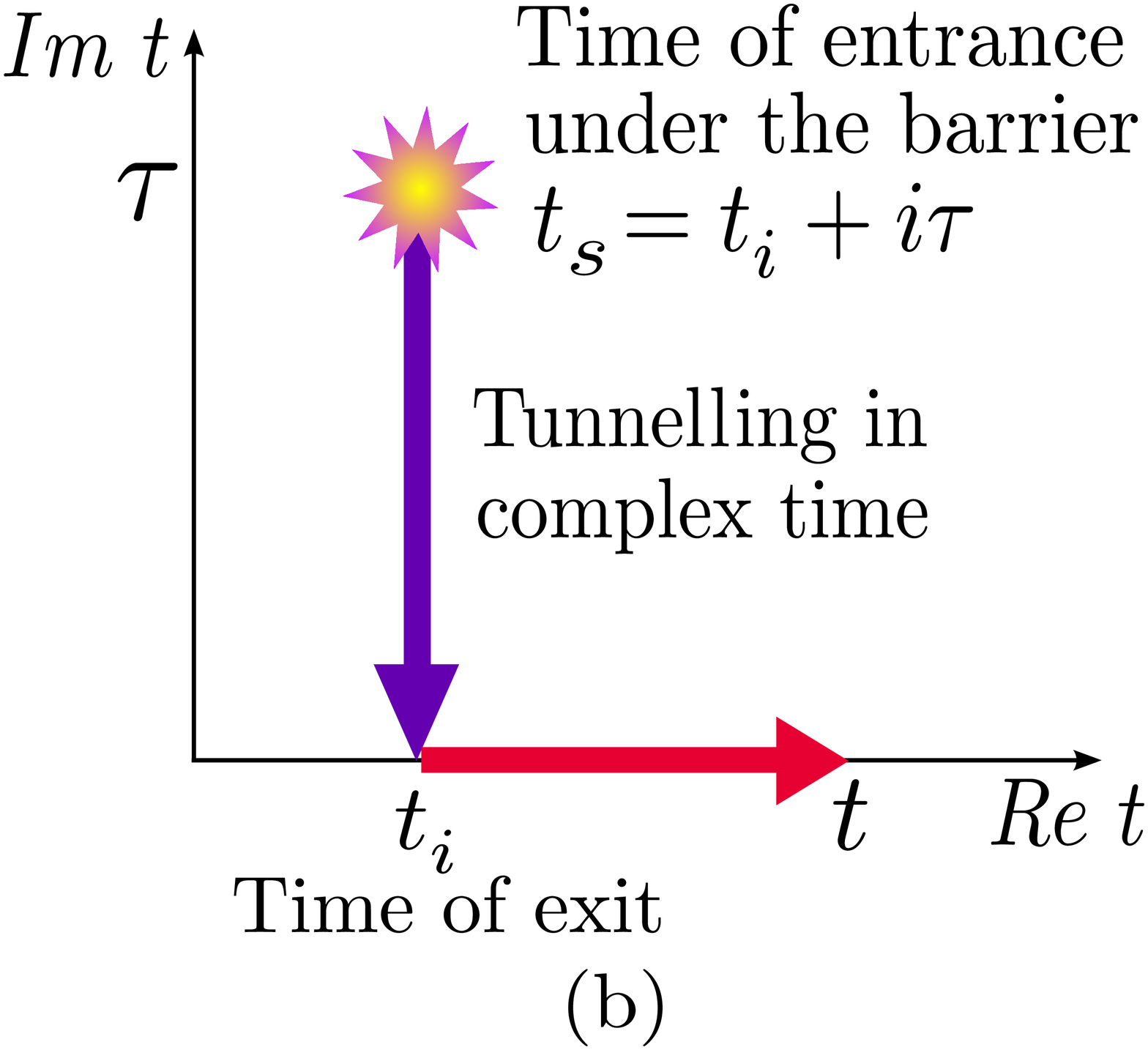}
\caption{(a) Schematic representation of the fundamental and control fields in multi-dimensional high harmonic spectroscopy. (b) Schematic representation of evolution in complex time $t_s=t_i+i\tau$. Quantum orbit enters the barrier at complex  time $t_s$, appears in continuum at ionization time $t_i$ and returns to the core at time t.}
\end{center}
\end{figure}

 High harmonic spectroscopy has been very successful in monitoring quantum orbits (also referred to as
 'quantum trajectories') \cite{Mairesse2,boutu},
  suggesting elegant schemes for measuring recombination \cite{Mairesse2,boutu,nirit1} and ionization times \cite{Shafir},
   turning the concept of quantum trajectories into a successful spectroscopic tool.
Here we propose and analyse the experimental scheme which allows for complete characterization of quantum orbits in complex systems by
detecting  the ionization delays,  imaginary ionization times (sub-cycle ionization rates),
and the recombination times  within the same high harmonic measurement.
In addition to the strong fundamental field ($\mathbf{E}(t)=-\partial \mathbf{A}/\partial t$) driving  the
ionization in the standard $1D$ HHG scheme
 \begin{eqnarray}
\label{eq:A3D}
\mathbf{A}_{1D}(t)=&-A_0\sin(\omega t)\,\mathbf{e}_z,
\end{eqnarray}
we introduce perturbative multicolour field
 \begin{eqnarray}
\label{eq:A3D}
\mathbf{A}_{\perp}(t,\phi_2,\phi_3)=&-\epsilon_2 A_0\sin(2\omega t+\phi_2)\,\mathbf{e}_x-
\epsilon_3 A_0\sin(3\omega t+\phi_2+\phi_3)\,\mathbf{e}_x,
\end{eqnarray}
combining the  second and the third harmonics of the  fundamental field.
The multicolour field is polarized orthogonally to the  linearly polarized fundamental field,  $\epsilon_2$ and $\epsilon_3$
 are small parameters, controlling the strength of the multicolour field.

Perturbation of the high harmonic signal
 induced by the multicolour field  is controlled via the two phase delays $\phi_2$ and $\phi_3$, where
  $\phi_2$ is the delay  between the fundamental field and the multicolour field and $\phi_3$ is the delay between the
   second and the third harmonic of the multicolour field itself.
The measurement of quantum orbits  is based on the optimization of  the harmonic signal as a function of the
harmonic number and the two phase delays $\phi_2$ and $\phi_3$,
   giving rise to the $3D$ high harmonic spectroscopy.

Our scheme extends the concept of the two-dimensional measurement suggested in \cite{Shafir}.
Setting $\epsilon_3=0$ in (\ref{eq:A3D}) one recovers the two-dimensional scheme of \cite{Shafir},
 in which  the fundamental field drives the strong field ionization while the
perturbative orthogonally polarized second harmonic field modulates the harmonic signal.
The two-dimensional scheme \cite{Shafir} allows one to reconstruct real ionization and recombination times,
 but the imaginary ionization times and thus the sub-cycle ionization rates remain hidden in these measurements.
 We show how and why extending the dimensionality of the measurement allows one to overcome this problem.

The $3D$ dipole $\D^{(mn)}_{3D}(t)$, which corresponds to leaving the ion in the state $|n^{(N-1)}\rangle$
after ionization and then recombination with the ion in the state
$|m^{(N-1)}\rangle$,  can be written as an integral over all ionization times $t'$ and all possible intermediate electron momenta $\p$ \cite{chapter}:
\bea
    \label{eq:dipmV31}
&&\D^{(mn)}_{3D}(t)= i \int_{0}^{t} dt'\int d\p \, \d_m^{*}(\p+\A_{3D}(t))\,a_{mn}(t,t')\,e^{-i\Sigma^n_{3D}(\p,t,t')}\times \nonumber \\
&&\hspace*{1.5cm}\times \Upsilon_n(\p+\A_{3D}(t')),\\
    \nonumber
  &&  \Upsilon_n(\p+\A_{3D}(t))=\left[\frac{\left[\p+\A_{3D}(t) \right]^2}{2} +I_{p,n}\right] \langle \p+\A_{3D}(t)|\Psi^{D}_n\rangle,
    \label{eq:Y_n}
     \nonumber \\ \label{eq:Sigma3D}
     && \Sigma^n_{3D}(\p,t,t')=S^n_{3D}(\p,t,t')+G^{n}_{3D}(\p,t,t'),\\
  \label{eq:S3D}
   &&  S^n_{3D}(\p,t,t')=\frac{1}{2}\int_{t'}^{t} [\p+\A_{3D}(\tau)]^2 d\tau+I_{p,n}(t-t'), \\
\label{eq:G3D}
&& G^{n}_{3D}(\p,t,t')=\int_{t'}^t d\tau U[\rmbf(\tau)],\qquad
 \rmbf(\tau)=\rmbf+\int_{t'}^{\tau} dt'' [\p+\A(t'')]. \nonumber
       \eea
Here $I_{p,n}$ is the ionization potential correlated to the electronic state $n$ of the ion. The amplitude $a_{mn}(t,t')$
  describes the laser-induced transitions in the ion between ionization and recombination \cite{chapter},
 $ \Upsilon_n(\p+\A_{3D}(t))$ describes the angular dependence of ionization, reflecting the structure of the Dyson orbital $|\Psi^{D}_n\rangle$.
 $\Sigma^n_{3D}(\p,t,t')$ is the electron action, which includes the  interaction of the departing electron with the core. This interaction is   encoded in $G^n_{3D}(\p,t,t')$ \cite{lisa2,jivesh}, where $U(\mathbf{r})$ is the core potential.
   Specific values of  $t'=t'_s$ and $\p=\p_s$, which minimize the action $\Sigma^n_{3D}(\p,t,t')$, define quantum orbits.
   In the simplest case when the effect of the core potential on the strong -field driven continuum electron is neglected, these trajectories are described by  $\mathbf{r}_s=\int_{t_s}^{t}d\xi(\mathbf{p}_{s}+\mathbf{A}_{3D}(\xi))$, but the effect of the core potential need not to be neglected, see \cite{PPTnew,lisa1,lisa2,jivesh}. These electron trajectories are launched at complex time $t_s=t_i+i \tau$ \cite{Keldysh,PPT}, when electron "enters" the tunnelling barrier, exiting the barrier at real time $t_i$ and  returning to the core at time $t$ (figure 1(b)). Finally,  $\mathbf{A}_{3D}(t,\phi_2,\phi_3)=A_{1D}(t)\,\mathbf{e}_z+A_{\perp}(t,\phi_2,\phi_3)\,\mathbf{e}_x$. The component $S^n_{3D}(\p,t,t')$ is the dominant part of the action associated with the laser-driven  electron dynamics.
 It is convenient to rewrite it by separating the motion in longitudinal and transverse directions:
 \bea
 &&\Sigma^n_{3D}(\p,t,t')=\Sigma^n_{1D}(p_{||},t,t')+\sigma_{2D}(p_{\perp},t,t',\phi_2,\phi_3),\\
  \label{eq:S1D} 
 && \Sigma^n_{1D}(p_{||},t,t')=S^n_{1D}(p_{||},t,t')+G^{n}_{1D}(\p,t,t'),\\
 &&  S^n_{1D}(p_{||},t,t')=\frac{1}{2}\int_{t'}^{t} [p_{||}+A_{1D}(\xi)]^2 d\xi+I_{p,n}(t-t'),\\
  &&  \sigma_{2D}(p_{\perp},t,t',\phi_2,\phi_3)=\frac{1}{2}\int_{t'}^{t} [p_{\perp}+A_{\perp}(\xi)]^2 d\xi.
    \label{eq:sigma2D}
    \eea
Consider the effect of the control field on the single channel harmonic dipole $\D_{3D}(t)\equiv\D^{(nn)}_{3D}(t)$, index of the channel will be omitted to simplify the notations.
 Due to the perturbative nature of orthogonally polarized field, the  harmonic dipole $\D_{3D}(t)$ formed after one half-cycle of the fundamental field can be written in the following form:
 \bea
    \label{eq:dipt}
    && \D_{3D}(t)=\left[\D_{1D}(t)+\mathcal{O}(\epsilon_{2,3})\right]e^{-i\sigma_{2D}(p_{s\perp},t,t'_s,\phi_2,\phi_3)}+c.c.,\\
  &&  p_{s\perp}(t,t'_s)=\frac{-1}{t-t'_s}\int_{t_s}^{t}A_{\perp}(\xi,\phi_2,\phi_3)d\xi.
  \label{eq:ps}
\eea
Equation (\ref{eq:dipt}) shows that  the full dipole factorizes into  the unperturbed dipole $\D_{1D}(t)$
 due to fundamental field and  perturbation $e^{-i\sigma_{2D}(p_{s\perp},t,t_s,\phi_2,\phi_3)}$ due to the control field.
 Mathematically, factorization is achieved by applying  the saddle point method \cite{chapter} to the integrals over $t'$ and $p_{\perp}$. In equation (\ref{eq:dipt}) $t_s$ is the ionization time for the $1D$-dipole $\D_{1D}(t)$, arising in the standard $1D$ HHG set-up involving only the fundamental field.
 The perturbative nature of the control field allowed us to  perform the Taylor expansion of $d_m^{*}(\p+\A_{3D}(t))$, $ G^n_{3D}(\p,t,t_s)$ and $\Upsilon_n(\p+\A_{3D}(t_s))$ around $\epsilon_{2,3}=0$, where $\D_{1D}(t)$ represents  the leading term of such expansion.
 The term $e^{-i\sigma_{2D}(p_{s\perp},t,t'_s,\phi_2,\phi_3)}$ represents the effect of the weak control field calculated along the unperturbed
  quantum trajectory. This is the leading term. The higher order terms, collected in $\mathcal{O}(\epsilon_{2,3})$, represent the effect of the control field on the trajectory itself, such as the modification of the ionization and recombination times.
 These higher order terms describe weaker processes, such as the generation of even harmonics due to symmetry breaking by the control field.
 Here we focus on the analysis of the dominant signal, i.e. that of the odd harmonics and therefore omit these terms. The validity of this approximation was verified by evaluating the effects of the perturbative control field on the ionization times. For the control field at $1-2\%$ of the driving field, the changes in real and imaginary ionization times also do not exceed the level of $2\%$. However, the effect of the control field on the electron motion in both real and imaginary times, 
 in particular on the accumulated action, is fully included in our analysis.


 The harmonic spectrum is given by the Fourier transform of the harmonic dipole collected over one or several cycles of the fundamental field:
   \bea
    \label{eq:dipw}
    && I_{3D}(N\omega)=\left|\int_{-\infty}^{\infty}dt  \D_{3D}(t)e^{iN\omega t}\right|^2.
\eea
  Since the unperturbed part $\D_{1D}(t)e^{iN\omega t}$   is a highly oscillating function of time $t$ and the perturbation  $e^{-i\sigma_{2D}(p_{s\perp},t,t_s,\phi_2,\phi_3)}$  is a slow function of time (equal to unity for $\epsilon_{2,3}\rightarrow 0$), the
  integral in (\ref{eq:dipw}) can be rewritten as:
 \bea
    \label{eq:dipw1}
    && I_{3D}(N\omega)=\left|\int_{-\infty}^{\infty}dte^{-i\sigma_{2D}(p_{s\perp},t,t_s,\phi_2,\phi_3)}\D_{1D}(t)e^{iN\omega t}\right|^2
    \simeq
    \nonumber \\
    &&\hspace{3cm}  \simeq Q^q_2(t_{rN},t_s,\phi_2,\phi_3) I_{1D}(N\omega),\\
    &&Q^q_2(t_{rN},t_s,\phi_2,\phi_3)=|e^{-i\sigma_{2D}(p_{s\perp},t_{rN},t_s,\phi_2,\phi_3)}|^2,
    \label{eq:dateQ2}
\eea
where $t_{rN}$ are the recombination times, or the times when the highly oscillatory phase  $\Sigma_{1D}(\p,t,t')+N\omega t$ becomes stationary,
$I_{1D}(N\omega)$ is the unperturbed harmonic signal. The action $\Sigma_{1D}(\p,t,t')$ is given by Eq.( \ref{eq:Sigma3D}), where $\epsilon_2=\epsilon_3=0$.

The harmonic spectrum is modulated by the multicolour field and the amplitude of modulation depends on the
delays $\phi_2$ and $\phi_3$.
The modulation is controlled by the gate function $Q^q_2(t_{rN},t_s,\phi_2,\phi_3)$.
The maxima of this function correspond to optimization of both ionization and electron return to the core in the multicolour field.
 The return is optimized when the electron lateral velocity at the time of ionization $t_i$ required for return at time $t_{rN}$ is equal to zero, $v_{\perp}(t_i,\phi_2,\phi_3)=0$ \cite{hadas}.
 It means that the probability of an electron to return to the ion is higher if it doesn't receive any additional
  lateral drift velocity $v_{\perp}(t_i,\phi_2,\phi_3)=p_{s\perp}(t_i,t_s)+A_{\perp}(t_i,\phi_2,\phi_3)\approx A_{\perp}(t_i,\phi_2,\phi_3)$ from the multicolour field when it exits the barrier at time $t_i$.
  Thus, the optimal phases are naturally found in the vicinity of the phases $\phi_2^0$, $\phi_3^0$, such that $A_{\perp}(t_i,\phi_2^0,\phi_3^0)=0$ (see also figures 2, 3).

 The width and the shape of the gate  $Q^q_2(t_{rN},t'_s,\phi_2,\phi_3)$ is determined by the lateral momentum distribution of the electron wave-packet formed during ionization.
 Since the method operates on the optimization of the harmonic signal perturbed by the lateral displacement of the electron,
the modulation contrast depends on how wide the electron wave-packet is in the lateral direction. The reconstruction relies on the explicit knowledge of the gate $Q^q_2(t_{rN},t'_s,\phi_2,\phi_3)$.
We stress, however, that the knowledge of the gate does not rely on the strong field approximation, but is due to the perturbative nature of the control field \cite{nirit1,nirit2}.
\begin{figure}
\label{figure2}
\begin{center}
\includegraphics[width=5 cm]{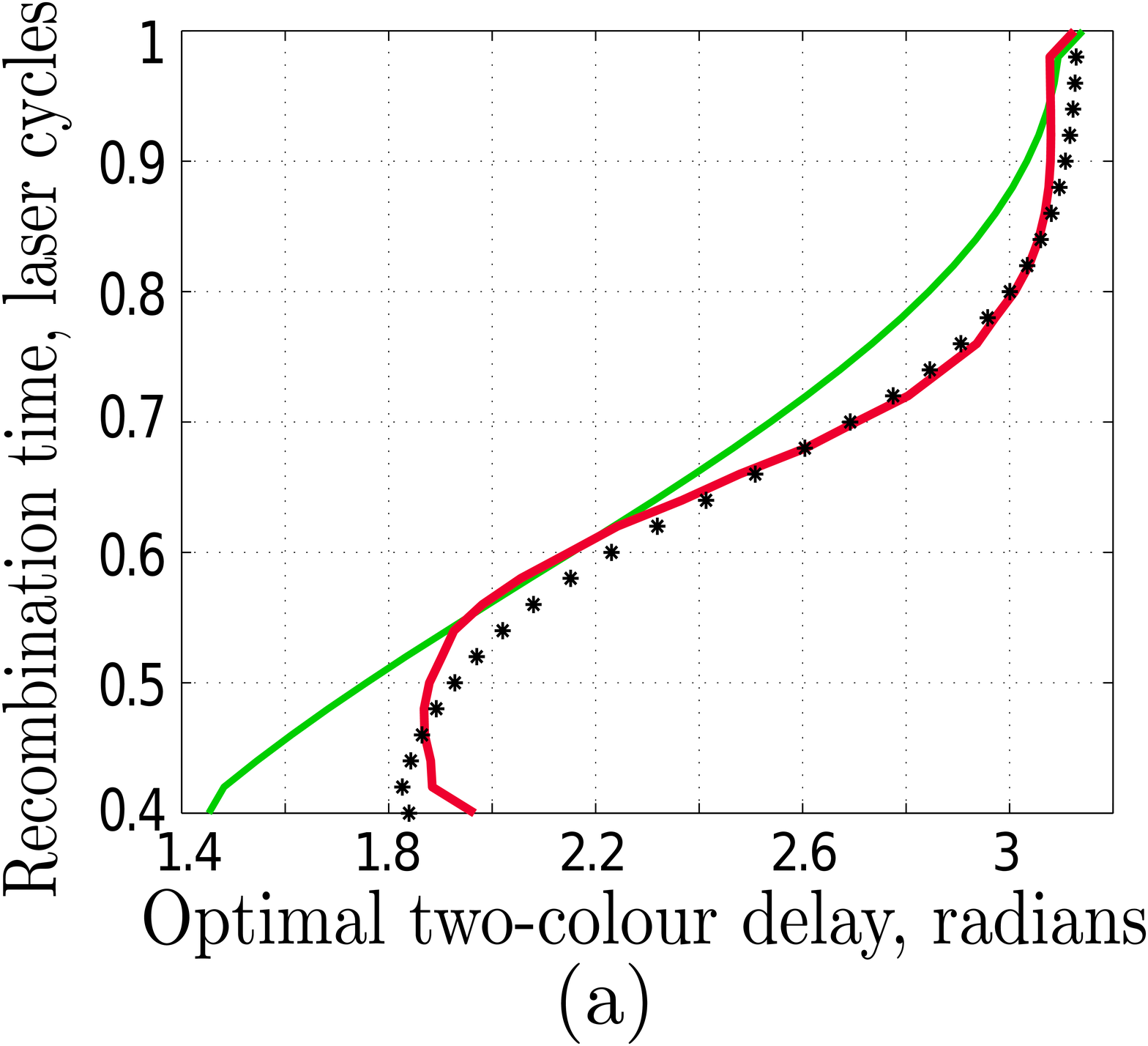}
\includegraphics[width=10 cm]{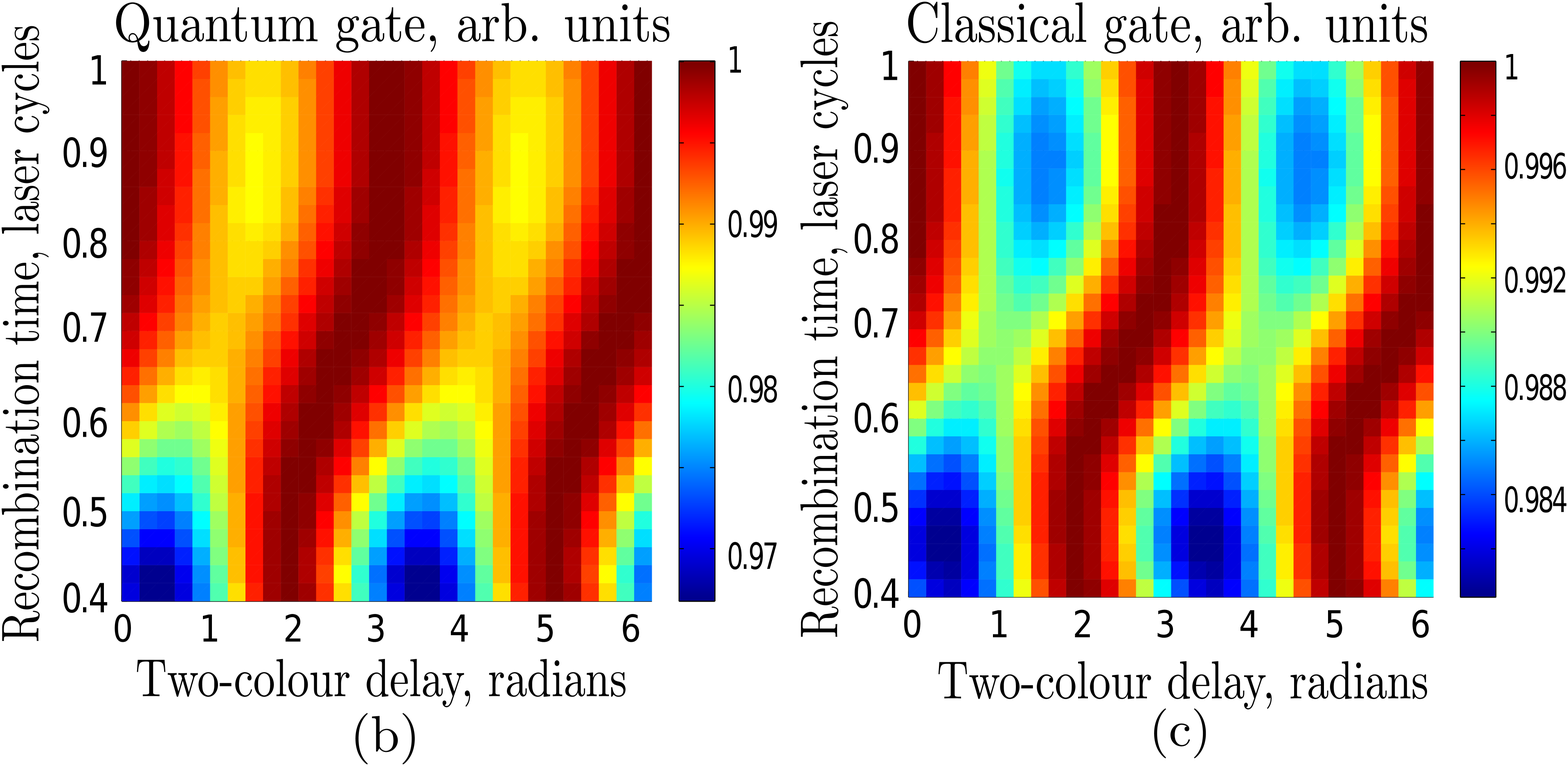}
\caption{Comparison of quantum and classical gates in $2D$ case:   $I_p=24.59$ eV, $\epsilon_2=0.07$, $\epsilon_3=0$, $I=1.36\times10^{14}$ W/cm$^2$, $\lambda=1600$ nm. (a) Two-colour delay (black dots) corresponding to the maximum of the quantum gate $Q^{q}_1$, two-colour delay (red triangles) corresponding to the  maximum  of classical gate $Q^{c}_1$, two-colour delay $\phi_0$ (green solid line) corresponding to zero of the  vector potential of the control field at the moment of ionization $t_i$.    (b) Contrast of modulation for the quantum gate, normalized to its maximum for each recombination time,
(c) Contrast of modulation for the classical gate.}
\end{center}
\end{figure}

Optimization of even harmonics  \cite{Shafir}  allows one to reconstruct  recombination times.
The subscript $N$ in $t_{rN}$ emphasises the connection between the recombination time $t_{rN}$ and the harmonic number $N$,
 which can be established through the analysis of even harmonics  \cite{Shafir} or independently using the RABBITT scheme \cite{boutu}.
Here we focus on the analysis of the dominant signal of odd harmonic  given by the first term in (\ref{eq:dipw1}),
 assuming that the recombination times are known or have been reconstructed, e.g.  using the  procedure developed  in \cite{Shafir}.
 Note, that for two-color control field the analysis of even harmonics developed in \cite{Shafir} can be extended
 to reconstruct not only the recombination phases, but also possible transverse momentum kick that the electron might have received
 due to momentum exchange with the core during ionization.
For the analysis of the odd harmonics we are left with
two unknowns: the real $t_i$ and the imaginary $\tau$ ionization times: $t_s=t_i+i \tau$.
Optimization of the $3D$ high harmonic signal provides two constraints, from which these two unknowns can be reconstructed:
\bea
\label{eq1}
\frac{\partial Q_2^q(t_{rN},t_s,\phi_2,\phi_3)}{\partial \phi_2}\vert_{\phi_2^{opt}(N)}=0,
\eea
\bea
\label{eq2}
\frac{\partial Q_2^q(t_{rN},t_s,\phi_2,\phi_3)}{\partial \phi_3}\vert_{\phi_3^{opt}(N)}=0.
\eea
Formally, these equations provide all necessary and sufficient conditions to reconstruct the two unknowns from the two equations.
Measurement of optimal phases $\phi_2^{opt}(N)$,  $\phi_3^{opt}(N)$ allows one to obtain $t_i(N)$ and $\tau(N)$ by solving  equations (\ref{eq1}) and(\ref{eq2}) for the gate given by (\ref{eq:dateQ2}).

An elegant way of solving these equations is revealed by the analysis of the $2D$ approach \cite{Shafir}. 
In the $2D$ approach, the gate $Q_1^q(t_{rN},t_i,\tau,\phi_1)$ is obtained by setting $\epsilon_3=0$ in (\ref{eq:dateQ2}).
It contains two unknowns $t_i(N)$ and $\tau(N)$,
 but it has only  one parameter at its disposal -- the optimal phase $\phi_2^{opt}(N)$ and, thus, only one constraint leading to  one equation.
  The reconstruction of $t_i(N)$ is possible due to the specific degeneracy of the $2D$ gate, which makes the optimal phase $\phi_2^{opt}(N)$ virtually insensitive to the imaginary time $\tau$.
To demonstrate this fact, we recall the simple man's picture of the "classical" gate \cite{Shafir}.
It results from ignoring the imaginary time in  the saddle point solution (\ref{eq:ps})
and disregarding  the effect of the control field on ionization. Mathematically, it corresponds to representing the gate by the
lateral electron velocity distribution after tunnelling in the fundamental field:
 \begin{eqnarray}
\label{eq:class}
&&Q_1^{c}(t_{rN},t_i,\phi_2)=e^{-\left[p_{c\perp}+A_{\perp}(t_i,\phi_2)\right]^2 \tau},\\
&&  p_{c\perp}(t,t_i)=\frac{-1}{t-t_i}\int_{t_i}^{t}A_{\perp}(\xi,\phi_2)d\xi.
  \label{eq:pscl}
\end{eqnarray}
Note that the imaginary time $\tau$ only determines the widths of the classical gate, but it does not influence the position of its maximum.
The return is optimal when the initial electron velocity in lateral direction is zero $v_{\perp}(t_i,\phi_2)\approx A_{\perp}(t_i,\phi_2)=0$ \cite{hadas}.

 By forcing the electron to exit the barrier  around the  zeros of the vector potential $A_{\perp}(t_i,\phi_2^{0}(N))\approx 0$
  one automatically suppresses the  effect of the control field on ionization, minimizing the difference between the classical and the quantum gates.
   Indeed, using (\ref{eq:dateQ2}) with $\epsilon_3=0$
 we obtain the $2D$ quantum gate $Q_1^q(t_{rN},t_s,\phi_2)$:
 \bea
   &&\sigma_{1D}(p_{s\perp},t,t_s,\phi_2)=\frac{1}{2}\int_{t_s}^{t} [p_{s,\perp}+A_{\perp}(\xi)]^2 d\xi,\\
   && p_{s\perp}(t,t_s)=\frac{-1}{t-t_s}\int_{t_s}^{t}A_{\perp}(\xi,\phi_2)d\xi,\\
  && Q_1^q(t_{rN},t_s,\phi_2)=e^{-2 \textrm{Im}\sigma_{1D}(p_{\perp},t,t_s,\phi_2)},
   \label{eq:quant1D}
\eea
 which can be further
 re-written in the following form:
    \label{eq:sigma2D}
 \begin{eqnarray}
\label{eq:quant}
&&Q_1^q(t_{rN},t_s,\phi_2)=e^{-\left[ p_{s\perp}-p_{i\perp}\right]^2\tau}e^{ -p^{2}_{i\perp}\tau f(\omega\tau)},\\
\label{eq:pionopt}
&&p_{i\perp}=-A_{\perp}(t_i,\phi_2)\frac{\sinh(2\omega\tau)}{2\omega\tau},\\
&&f(\omega\tau)=\omega\tau \sinh(4\omega\tau)/\sinh^2(2\omega\tau)-1.
  \label{eq:pion}
\end{eqnarray}
The physical meaning of  $p_{i\perp}$ given by (\ref{eq:pionopt}) is that it optimises ionization.
 Optimization of ionization is achieved for those $p_{i\perp}$, which solve the equation: $\partial \textrm{Im} \sigma_{1D}(p_{\perp},t,t_s,\phi_2)/\partial p_{\perp}=0$.
 Note that  $ p_{s\perp}$ is a solution of the equation $\partial  \sigma_{1D}(p_{\perp},t,t_s,\phi_2)/\partial p_{\perp} =0$, which optimizes the electron return.
 In deriving (\ref{eq:quant}) we have neglected a small imaginary component of $ p_{s\perp}$, which does not affect our results or conclusions.
 Thus, (\ref{eq:quant}) shows that the maxima of the quantum gate $Q_1^q(t_{rN},t_s,\phi_2)$ are indeed determined by the interplay between the optimal  ionization and return conditions, while the classical gate (\ref{eq:class}) only optimizes the electron return.
However, substituting $A_{\perp}(t_i,\phi_2^{0})\simeq0$ yields
 \begin{eqnarray}
\label{eq:class_red}
&&Q_2^q(t_{rN},t_s,\phi_2)\simeq e^{- p^2_{s\perp}\tau},\\
 \label{eq:psred}
&& p_{s\perp}( \phi_2)=-\alpha\left[\cos(2\omega t_r+\phi_2)-\cosh(2\omega \tau)\cos(2\omega t_i +\phi_2)\right],\\
\label{eq:quant_red}
&&Q_2^{c}(t_{rN},t_i,\phi_2)=e^{-p^2_{c\perp}\tau},\\
&& p_{c\perp}( \phi_2)=-\alpha\left[\cos(2\omega t_r+\phi_2)-\cos(2\omega t_i +\phi_2)\right],
  \label{eq:pcred}
\end{eqnarray}
where $\alpha=\epsilon_2A_0/2\omega(t_{rN}-t_i)$.
Thus, the classical and quantum gates are virtually identical around $A_{\perp}(t_i,\phi_2^{0})\simeq0$, since in this case  $p_{s\perp}\approx p_{c\perp}$.
Indeed, the modulation of the signal is dominated by the first term in (\ref{eq:psred}), (\ref{eq:pcred}), because the second term reaches its maximum $\cos(2\omega t_i +\phi_2)\simeq 1$ around $A_{\perp}(t_i,\phi_2^{0})\simeq0$ and thus changes weakly as we vary $\phi_2$ in the vicinity of $\phi_2^{0}$. The difference between the quantum and the classical gates always increases in the non-adiabatic regime, when $\omega\tau \gg 1$.

Therefore, if the harmonic signal maximises when the electron exits the barrier around the zero of the vector potential
 of the control field $A_{\perp}(t_i,\phi_2^{0})\simeq0$, the maxima of the classical and quantum gates are
  very close and both gates are virtually indistinguishable (see figure 2(a)).
  The only difference between the two is in the contrast of the modulation (figure 2(b)) introduced by the control field.
Thus, the classical gate can be used to reconstruct  real ionization time $t_i$ -  the time when electron exits the barrier \cite{Shafir}
without any apriori knowledge about the imaginary ionization time.
 The   information about the imaginary time is encoded in the contrast of the modulation.
However, in the experiment it is very challenging to use the modulation contrast as a reconstruction variable, since it is obscured by the measurement noise.

In the $2D$ set-up optimization of ionization and recombination for odd harmonics occurs for the same phase delays,
 with much stronger contrast for recombination. 
A qualitative interpretation of why the classical gate in the $2D$ scheme is successful is as follows.
Tunnelling tends to confine the electron motion
 in the lateral direction, forcing the electron to move along the polarization direction of the laser field, where the barrier is thinner.
 Thus, loosely speaking, the sub-barrier part of the electron trajectory is much less affected by the control field acting in the lateral direction, than the continuum part of the trajectory,
 keeping us away from detecting imaginary times and manipulating sub-barrier trajectories.
 However, the sub-barrier part can be accessed with the additional control field.

Indeed, the degeneracy between the quantum and classical gates arising in the $2D$ case can be removed by increasing the dimensionality of the measurement.
In the $3D$ case, the  harmonic signal will again  maximize in the vicinity of the phase delays  $ \phi_2^{0}$ and $\phi_3^{0}$
  for which the total vector-potential given by (\ref{eq:A3D}) at time $t_i$
is equal to zero, $\mathbf{A}_{\perp}(t_i,\phi_2^{0},\phi_3^{0})\simeq 0$. Since the control field
now operates with  two colours $2\omega$ and $3\omega$, each of the respective components of $\mathbf{A}_{\perp}(t_i,\phi_2^{0},\phi_3^{0})$
 can be non-zero at the time $t_i$, while their sum is still equal to zero.
In this case, the degeneracy between the quantum and classical gates is removed, because such degeneracy requires that every component of the
 vector potential of the multicolour field is close to zero at time $t_i$.
In the $3D$ case one can therefore control the degeneracy by changing the phase $\phi_3$, using it do develop simple reconstruction schemes.
 For example, setting $\phi_3=0$ we recover the degeneracy in the $3D$ case and we can use the classical gate to
  reconstruct the the real ionization times $t_i$ and the recombination times $t_{rN}$ following the same procedure as in the $2D$ case \cite{Shafir}.
   Figure 2(a) shows that maxima of classical and quantum gates are indeed in excellent agreement in the $3D$ set-up for $\phi_3=0$, since
  the vector-potentials of $2\omega$ and $3\omega$ are both close to zero at time $t_i$.
 Selecting a cut through the $3D$ HHG spectrum, corresponding to  $\phi_3=0$ we can apply the classical reconstruction procedure \cite{Shafir}
 to retrieve real ionization times $t_i$ and (if we use even harmonic spectra) recombination times $t_{rN}$, without any prior knowledge about the imaginary times.
\begin{figure}
\label{figure3}
\begin{center}
\includegraphics[width=5 cm]{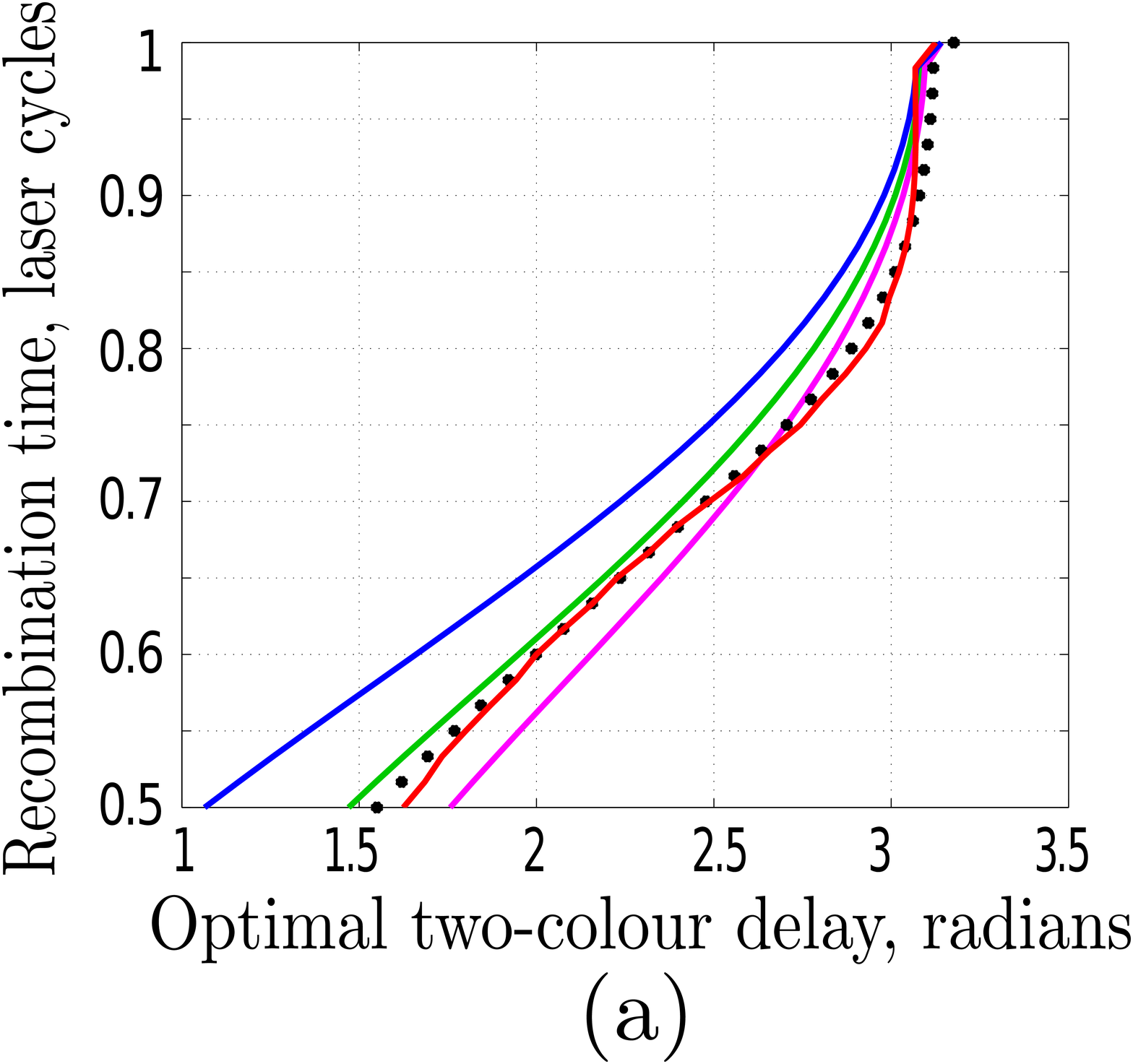}
\includegraphics[width=5 cm]{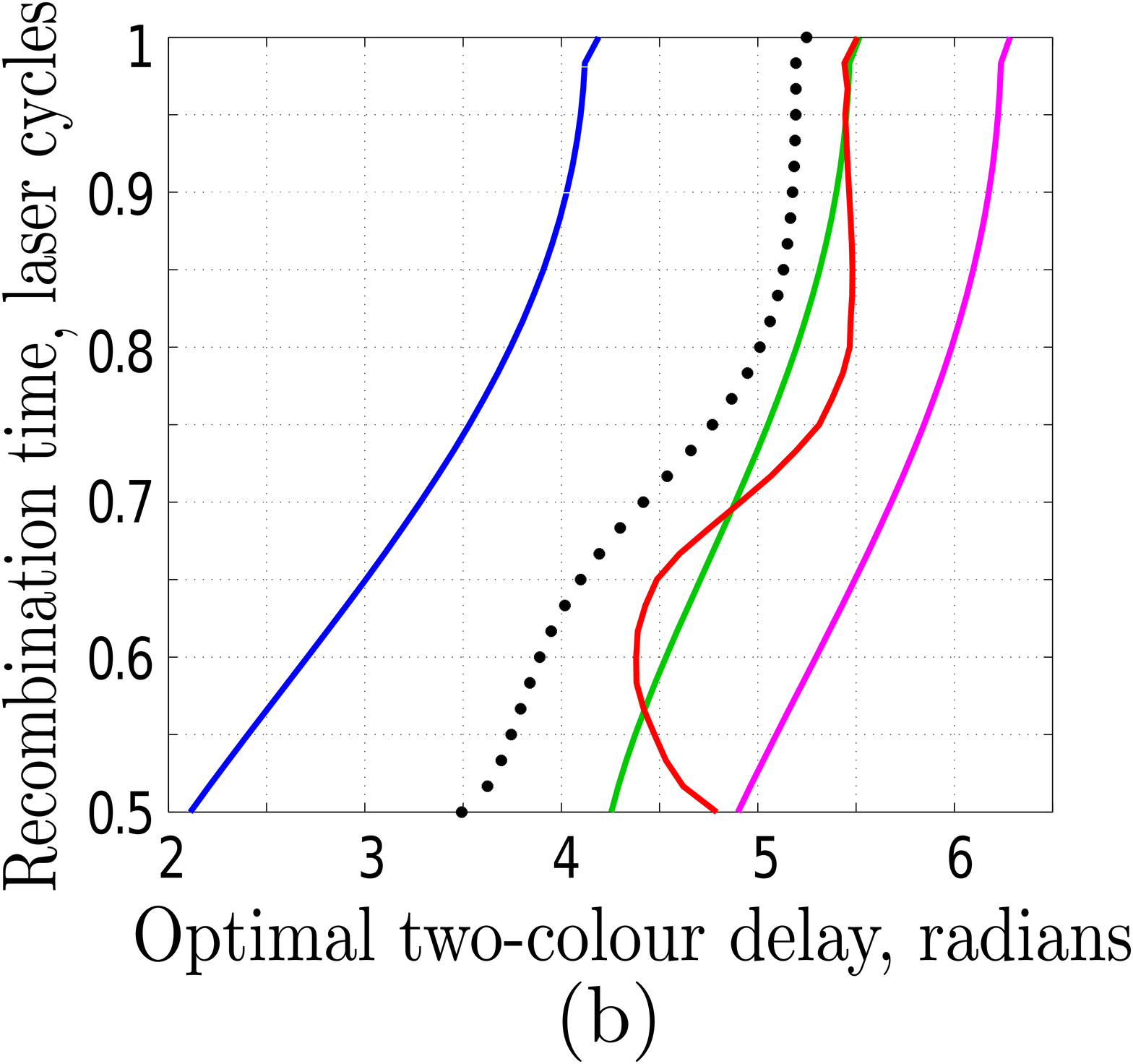}
\includegraphics[width=5 cm]{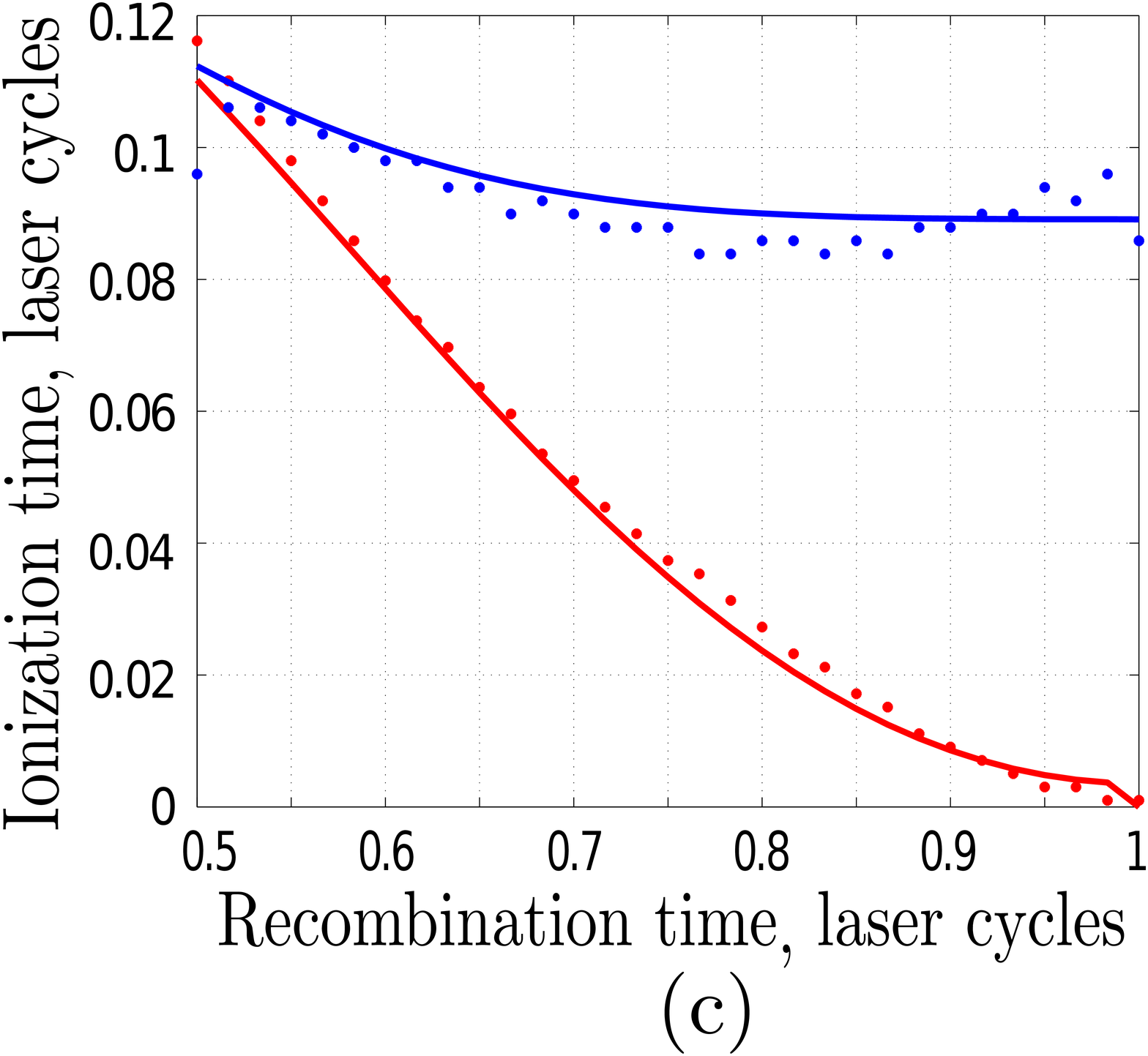}
\caption{Slice by slice reconstruction of the $3D$ HHG data: $I_p=24.59$ eV, $\epsilon_2=0.07$, $\epsilon_3=0.05$, $I=1.36\times10^{14}$
W/cm$^2$, $\lambda=1600$ nm. (a) Optimal delay $\phi_2^{opt,1}(N)$ in the degenerate case $\phi_3=0$   corresponding to the maximum of the quantum gate $Q^{q}_2$ (black dots); corresponding to the  maximum  of classical gate $Q^{c}_2$ (red);   corresponding to zero of the  vector potential of the control field at time $t_i$ (green);  corresponding to zero of the  vector potential of $2\omega$ field at time $t_i$ (magenta);  corresponding to zero of the  vector potential of $3\omega$ field at time $t_i$   (blue).
 (b) Optimal delay $\phi_2^{opt,2}(N)$ for non-degenerate case $\phi_3=2.1$ rad. The same notations are used. (c) Reconstruction of ionization time. Red dashed curve represent theoretical values of $t_i$,
  red dots - reconstructed values of ionization time $t_i$. Blue solid curve represent theoretical values of $\tau$,  blue triangles - reconstructed values of imaginary ionization time $\tau$.}
\end{center}
\end{figure}
Once we have reconstructed the real ionization times, we can remove the degeneracy by changing phase   $\phi_3$. Figure 2(b) shows that when $\phi_3\approx 2.1$ rad, the maxima of classical and quantum gates are  very different, since the vector potential of each individual colour is non-zero at the time $t_i$.  In this case the classical description breaks down, since 
the second and the third harmonics of the driving laser field act out of phase on the tunnelling electron, introducing non-adiabatic effects during ionization. As a result, the optimal lateral velocity during tunneling is no longer equal to zero: the electron receives a lateral "kick" from the control field during tunneling. While these effects still have perturbative nature in the sense that they do not change significantly the ionization times, they introduce the dependence of the  optimal delay on the dynamics under the barrier, specifically on the value of the imaginary ionization time. In the language of quantum orbits, in the $3D$ set-up one can strongly affect the trajectory both before and after the exit from the barrier, getting access to both real and imaginary times.
 Quantum gate depends on both real and imaginary times, but since the real time has been already reconstructed for $\phi_3=0$, we can now reconstruct the imaginary time using only one "slice" of the $3D$ spectrum, corresponding to $\phi_3=2.1$ rad. 
The accuracy of the "slice-by-slice" reconstruction maximizes for the minimal degeneracy between the quantum and classical gates. The degeneracy is controlled by the phase $\phi_3$ 
\footnote{Several values of $\phi_3$ in the range between $0$ and $2\pi$ were tested for the particular values of the strength of the fields, the ionization potential and the fundamental wavelength. The classical approach describes the position of the optimal delay within the error of $0.2$ rad
when $\phi_3$ is within the intervals $[0,0.8]$ rad and $[4.8,2\pi]$ rad. The interval $\phi_3\in[1.8,2.8]$ rad provides good accuracy for the 
reconstruction in the example considered below. In general, one does not have to use the "slice-by-slice" reconstruction; the reconstruction  can also 
be performed by choosing  two arbitrary values of $\phi_3$  in the interval $[-1.5,\pi]$ rad.}.

 We test the proposed reconstruction procedure using the simplest example of  the short-range electron-core interaction and the spherically symmetric ground state. This  problem can be solved exactly (see e.g. \cite{PPT}) and real and imaginary ionization times are well known for this case. We first simulated the $3D$ spectrum using ( \ref{eq:dipmV31}) and the method described in \cite{chapter} for the following set of parameters: $I_p=24.59$ eV, $\epsilon_2=0.07$, $\epsilon_3=0.05$, $I=1.36\times10^{14}$
W/cm$^2$, and the fundamental wavelength $\lambda=1600$ nm.
  In our test case $a_{nn}=1$, $G=0$ and  since the ground state is spherically symmetric the specific values of
  $\Upsilon(\p+\A_{3D}(t'))$ and $d^{*}(\p+\A_{3D}(t))$
 do not affect positions of the  maxima in the $3D$ harmonic spectrum. We first consider the slice of the spectrum corresponding to $\phi_3=0$
  and find the optimal phase delay  $\phi_2^{opt,1}(N)$, for which the harmonic signal maximises. Assuming that $t_{rN}$ are known, we substitute these values of $\phi_2^{opt,1}(N)$ into the
  optimal equation for the classical gate $\partial Q_1^{c}(t_{rN},t_i,\phi_2) / \partial \phi_2|_{\phi_2=\phi_2^{opt,1}(N)}=0$ to retrieve $t_i(N)$.
  The results of the reconstruction are in excellent agreement with the well-known theoretical values (figure 3 (a)).
  Now we take a slice of the $3D$ spectrum, corresponding to  $\phi_3=2.1$ rad and find the optimal phase delay  $\phi_2^{opt,2}(N)$ for which the  harmonic signal maximises.
  We use the quantum gate (\ref{eq:quant1D}) and substitute $\phi_2^{opt,2}(N)$ and $t_i(N)$ into the corresponding equation
    $\partial Q_1^{q}(t_{rN},t_i(N),\tau(N),\phi_2)/ \partial \phi_2|_{\phi_2=\phi_2^{opt,2}(N)}=0$ to retrieve $\tau(N)$.
     The results of reconstruction are in excellent agreement with the theoretical values (Fig. 3 (b)).
     Thus, the reconstruction of $3D$ spectrum can be as simple as it was in the original $2D$ proposal \cite{Shafir}.
     However, the $3D$ high harmonic spectroscopy should allow us to reconstruct real and imaginary ionization times in complex systems.
      We expect that these times will be significantly different from the ones used in the test-case example here. 
Note that the "slice-by-slice" reconstruction  decouples equations (\ref{eq1}) and (\ref{eq2}) using the properties of the classical gate. In general, the classical gate becomes less accurate in the non-adiabatic regime of strong-field ionization $\gamma\gg1$. In this case, the "slice by slice" reconstruction should be substituted by the more accurate "direct reconstruction", which  uses two arbitrary values of $\phi_3$ and solves the coupled equations (\ref{eq1}) and (\ref{eq2}) to find the two unknowns: the real  $t_i$ and imaginary  $\tau$ ionization times.

 The $3D$ high harmonic spectroscopy provides sufficient information for  full characterization of quantum orbits, detecting ionization and recombination times and initial velocity of the electron both in longitudinal and transversal direction. Here we focused on the analysis of the dominant -odd harmonic signal, which allows one to reconstruct real and imaginary ionization times.
 We have shown how one can perform "slice" by "slice" reconstruction of the $3D$ data by selecting specific values of $\phi_3$, keeping analysis as simple as in the $2D$ case, but adding the additional capability of reconstructing the imaginary ionization times.
 Our method is not limited to the analysis of odd harmonics.
 In the most general case, all parameters of quantum orbits: real and imaginary ionization times, the recombination times and transverse electron momentum can be reconstructed from four equations \footnote{These four equations are: equations (\ref{eq1}) and(\ref{eq2}) and two similar equations formulated for even harmonics.}, given by the optimization of the harmonic signal versus  the two phase delays $\phi_2$ and $\phi_3$  for  even and odd harmonics, while the initial velocity in longitudinal direction is uniquely determined by the reconstructed recombination and ionization times. The analysis of even harmonics has already been performed in the $2D$ set-up for He atom \cite{Shafir}. Similar analysis should be possible for linear molecules aligned at $0^o$ and $90^o$ to the fundamental field and other  geometries preserving central symmetry. 
  Thus, time delays, energy and momentum exchange, modification of ionization rate triggered by electron-hole interaction during ionization are directly linked  to modifications of quantum orbits and can be detected using multidimensional HHG, providing a complementary insight into attosecond dynamics of electron rearrangement upon ionization.
  \ack
  We gratefully acknowledge the  support of the German-Israeli Foundation (GIF 710883).
We thank M. Ivanov  for  useful suggestions and patience in weathering our complaints. We thank N. Dudovich and H. Soifer for  fruitful discussions.

  \end{document}